\documentclass[
 ,final            
  ]
  {aipproc}

\layoutstyle{8x11single}

\begin{document}
\bibliographystyle{aipproc} 
\title{Perturbative and Non-Perturbative Origins of the Proton Sea}


\author{Mary Alberg}{
  address={Department of Physics, Seattle University, Seattle, WA 98122, USA},
altaddress={Department of Physics, University of Washington, Seattle, WA 98195, USA} 
}
\author{Tyler Matossian}{
  address={Department of Physics, Seattle University, Seattle, WA 98122, USA} ,altaddress={Department of Chemistry, University of Southern California, Los Angeles, CA 90089, USA}}

\begin{abstract}
Deep Inelastic Scattering and Drell-Yan experiments have measured a light flavor asymmetry in the proton sea. The excess of $\bar{d}$ over $\bar{u}$ quarks can be understood in many models, but the ratio $\bar{d}(x)/\bar{u}(x)$ measured by Fermilab E866 has not been successfully described. Fermilab E-906 will probe the kinematic dependence of this ratio with better resolution and extend it to higher $x$. We have developed a hybrid model that includes both perturbative and non-perturbative contributions to the proton sea. A meson cloud formalism is used to represent the non-perturbative fluctuation of the proton into meson-baryon states. We include perturbative processes by using a statistical model that uses Fock states of quarks, antiquarks and gluons to represent the parton distributions of the `bare'  hadrons in the meson cloud.  We compare our results to the E866 data. \end{abstract}

\maketitle


\section{Introduction}

Asymmetry in the proton sea has been well-established by experiment \cite{NMC, NA51,E866}. Many theoretical models for the asymmetry have been proposed, based on Pauli-blocking, a meson cloud, chiral perturbation theory, instantons, or statistical processes.  For reviews, see Kumano \cite{Kumano98}, and Peng and Garvey \cite{P&G}. In this paper we examine both perturbative and non-perturbative contributions to the proton sea, and show that either will lead to an asymmetry. Since both contributions are present, we propose a hybrid model, and compare to experiment.
\section{Non-perturbative contribution: the Meson Cloud}
 As first suggested by Thomas \cite{Thomas83}, and later by Henley and Miller \cite{HM}, the meson cloud of the proton provides a natural explanation of the asymmetry. The fluctuation of the proton ($uud$) into a neutron ($udd$) and a $\pi^+$ ($u\bar{d}$) creates an excess of $\bar{d}$ over $\bar{u}$.
The meson cloud is represented in a Fock state expansion of the proton in terms of mesons and baryons
\begin{equation}
\mid p\rangle = \sqrt{Z}\mid p\rangle_{\rm bare} +
 \sum_{MB}\int dy\;
d^2\vec{k}_\perp\; \phi_{BM}(y,k_\perp^2)
\mid  B(y,\vec{k}_\perp) M(1-y, - \vec{k}_\perp)\rangle .
\end {equation}
The leading term in the expansion is the `bare' proton, taken to consist of valence quarks alone or valence quarks plus a symmetric light sea. Z is a normalization constant,  $\phi_{BM}(y,k_\perp^2) $
is the probability amplitude for finding a physical nucleon in a state
consisting of a baryon $B$ with longitudinal momentum fraction $y$ and meson
$M$ of 
momentum fraction $(1-y)$, each with squared transverse relative momentum $k_\perp^2$.
The quark distribution functions $q(x)$ are given by
\begin{equation}
q(x) = q_{\rm bare}(x) + \delta q(x) \; ,
\end{equation}
with $q_{\rm bare}(x)$ the quark distribution function of the `bare' proton, and 
\begin {eqnarray}
\delta q(x) = \sum_{MB}
\left(\int_x^1 f_{MB}(y) q_M \bigg(\frac{x}{y}\bigg)\frac{dy}{y} \;
+\int_x^1 f_{BM}(y) q_B\bigg(\frac{x}{y}\bigg)\frac{dy}{y}\right).
\end{eqnarray}
%
\begin{figure}[t]
  \centerline{
    \mbox{\includegraphics[width=2.50in,height=1.60in]{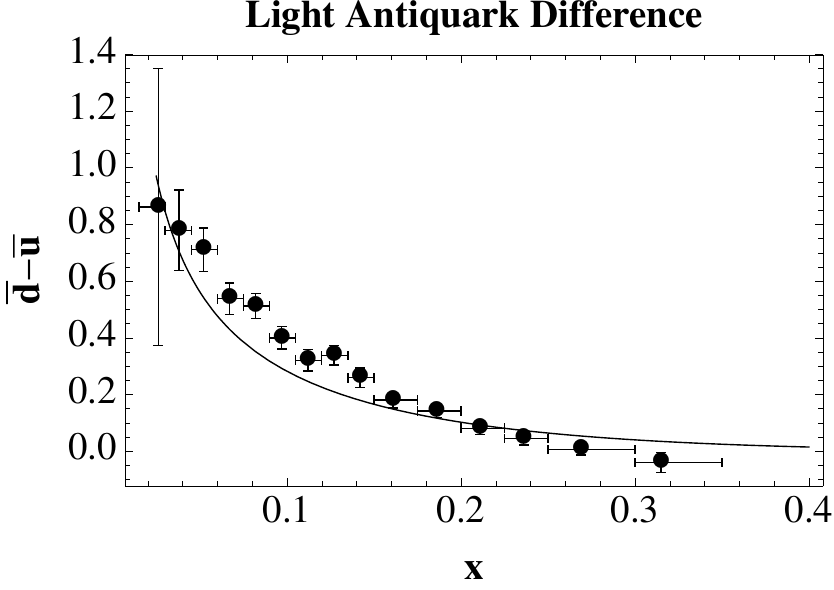}}
    \mbox{\includegraphics[width=2.50in,height=1.60in]{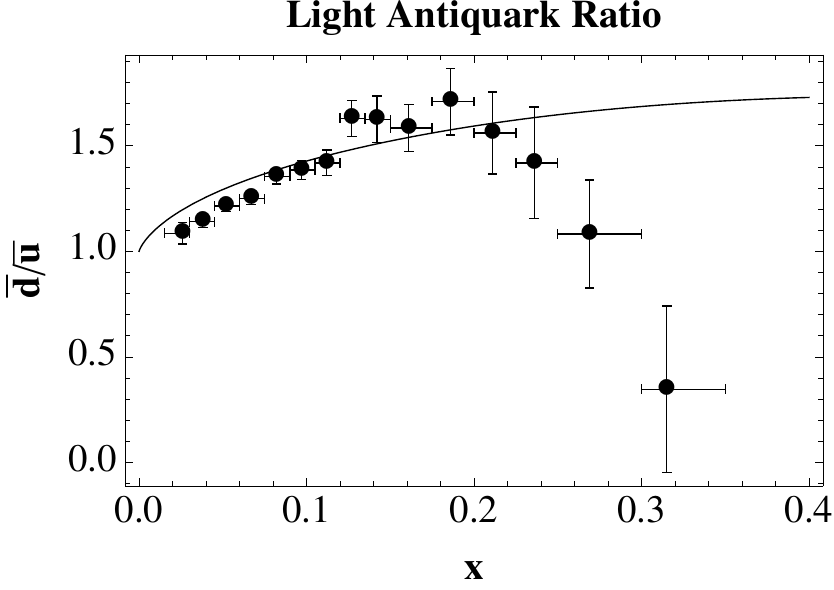}}
  }
  \caption{Meson cloud calculation for the proton sea asymmetry. Two cutoffs are used, $\Lambda_{NM}=0.92$ for nucleon-meson terms, and $\Lambda_{\Delta M}=0.98$ for Delta-meson terms. For these cutoffs, the 'bare' proton probability $Z = 0.58$,  $<N \pi> = 0.20$, $<\Delta \pi > = 0.11$, and higher order terms ($N\rho ,\Delta\rho,p\sigma,p\omega,p\eta$) represent 11\% of the proton. The data is from E866 \cite{E866}.}
  \label{MCM}
  \end{figure}
The splitting functions $ f_{MB}(y)$  depend on coupling constants, form factors, and cutoffs.
Models differ in the number of meson-baryon states included in the expansion, the parton distributions $q_M$ and $q_B$, form factors, cutoffs and coupling constants. The dominant terms in the expansion are $p \rightarrow n \pi^+$ and $p\rightarrow p \pi^0$. The first term causes a sea flavor asymmetry. The second term contributes equal numbers of $\bar{d}$ and $\bar{u}$, so it will affect the ratio $\bar{d}/\bar{u}$, but not the difference $(\bar{d} - \bar{u})$.  We have shown \cite{AHM} that the inclusion of the $\omega$ isoscalar contribution improves agreement with experiment, and Huang {\it et al.} \cite{Huang} find that an improved description of the data at higher $x$ is obtained if the $\sigma$  is included as well. However, the theoretical ratio still remains above 1 for large $x$.
A full meson cloud calculation should include all higher order terms in the expansion,  
$p \rightarrow \Delta \pi,  N \rho, \Delta \rho, p\, \omega, ...$ 
 In Fig.~\ref{MCM} we show our calculation with only 2 cutoff values, determined by a fit to the experimental integrated asymmetry  \cite{E866} of $0.118 \pm 0.012$ .  We can improve agreement with experiment by further adjustment of cutoffs, but the ratio remains above 1 for larger values of $x$. 
\section{Perturbative contribution:  statistical model}
Zhang {\it et al.}  \cite{ZZY} have used a simple statistical
model which is based on a Fock state expansion of the proton in terms of its quark and gluon states: 
\begin{equation}
|p> =\sum_{ijk} c_{ijk}|\{uud\}\{ijk\}>,
\end{equation}
with $i$ the number of $\bar{u}u$ pairs, $j$ the number 
of $\bar{d}d $ pairs and $k$ the number of gluons. 
 The probability of finding a proton in the state
$|\{uud\}\{ijk\}>$ is $\rho_{ijk}=|c_{ijk}|^2$.
 Taking into account the processes $q \leftrightarrow q\, g$, 
$g \leftrightarrow q\, \bar{q}$, and $g\leftrightarrow gg$,
detailed balance between any two states leads to
\begin{equation}
\frac{\rho_{ijk}}{\rho_{000}} = \frac{2}{i!(i+2)!j!(j+1)!}\prod_{n=0}^{k-1}\frac{3+2i+2j+n}{(3+2i+2j)(n+1)+\frac{n(n+1)}{2}},
\end{equation}
and an integrated asymmetry of 0.123, in remarkable agreement with experiment.
Parton distribution functions are determined by a Monte Carlo simulation of the distribution of momenta among the $n$ partons in each Fock state.
  \begin{figure}[b]
  \centerline{
    \mbox{\includegraphics[width=2.50in,height=1.60in]{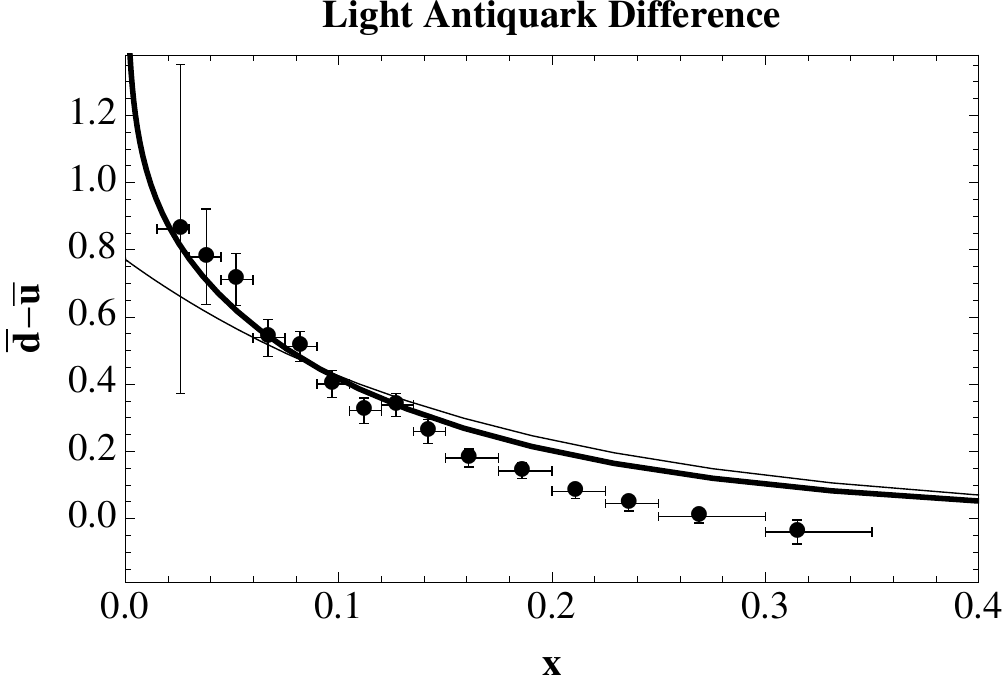}}
    \mbox{\includegraphics[width=2.50in,height=1.60in]{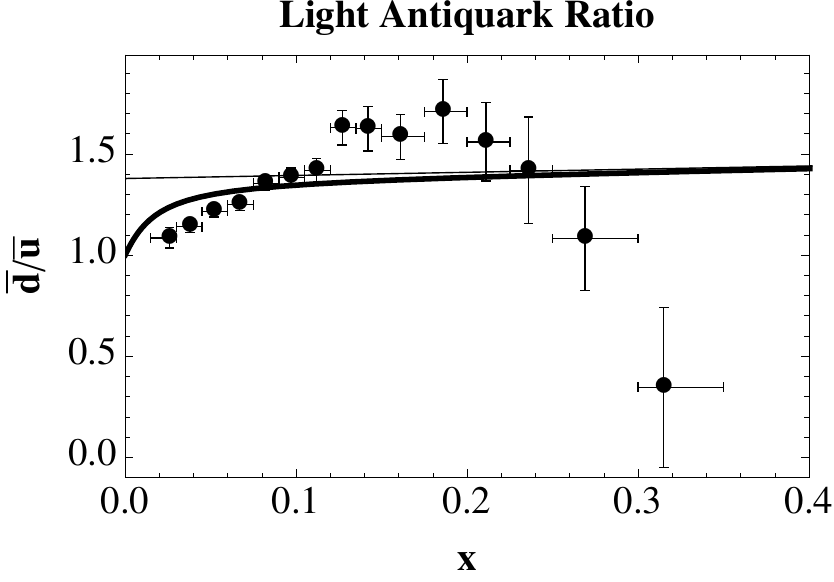}}}
  \caption{Statistical model calculation of the proton sea asymmetry. The thin line is calculated at the starting scale of the model, the thick line is calculated for parton distributions evolved to 54 GeV$^2$, the scale of the E866 experiment \cite{E866}.}
  \label{stat}
  \end{figure}
  In Fig.~\ref{stat} we show our calculation of the sea asymmetry, using this model at the starting scale and our evolution to the experimental scale. As for the meson cloud model, the difference is reasonable, but the ratio is relatively constant for larger $x$, in disagreement with experiment.
 \begin{figure}[t]
  \centerline{
    \mbox{\includegraphics[width=2.50in,height=1.60in]{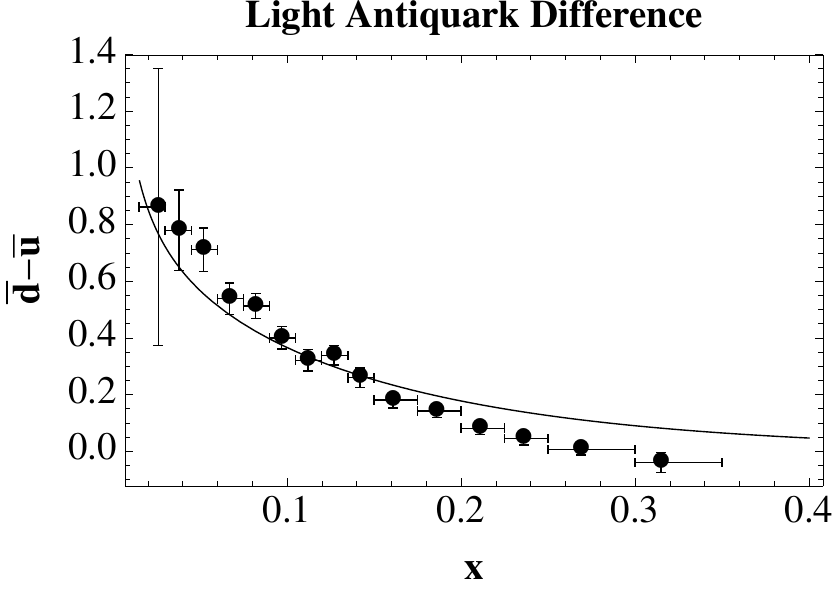}}
    \mbox{\includegraphics[width=2.50in,height=1.60in]{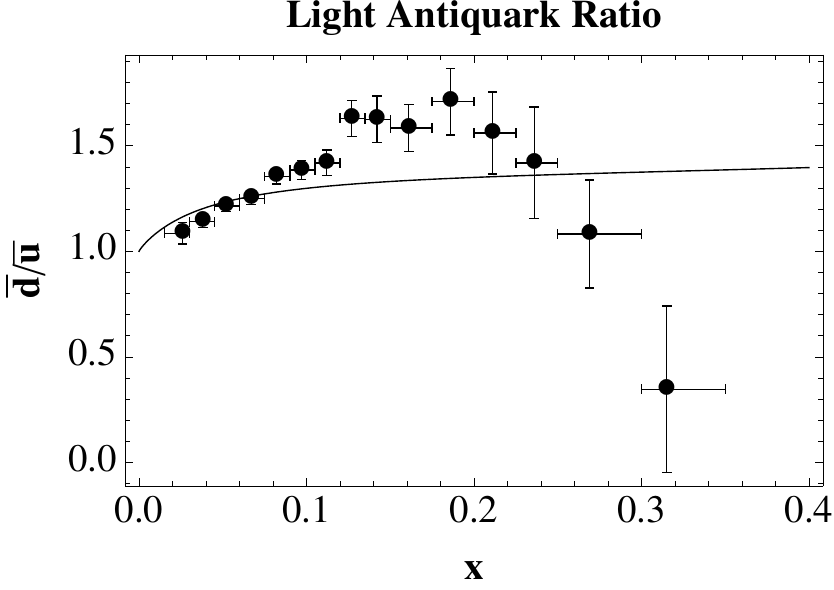}} }
  \caption{Preliminary hybrid model calculation for the proton sea asymmetry. Two cutoffs are used, $\Lambda_{NM}=0.83$ for nucleon-meson terms, and $\Lambda_{\Delta M}=1.05$ for  Delta-meson terms. For these cutoffs, the 'bare' proton probability $Z = 0.63$,  $<N \pi> = 0.15$, $<\Delta \pi > = 0.15$, and higher order terms ($\Delta\rho,N\rho ,p\sigma,p\omega,p\eta$) represent 8\% of the proton. The data is from E866 \cite{E866}.}
  \label{hybrid}
  \end{figure}
\section{A hybrid model}
We have proposed that since both perturbative and non-perturbative processes contribute to the proton sea, a hybrid model should be used to incorporate both mechanisms. We use the statistical model to represent the the `bare' proton parton distribution in the meson cloud model, which introduces an additional sea asymmetry. Our preliminary results are shown in Fig~\ref{hybrid}. Only 2 cutoffs were used, determined by a fit to the integrated asymmetry. The difference is well-fit, but the ratio is too flat. Calculations with further variation of the meson-baryon cutoff values are in progress \cite{AMhybrid}.  
\section{Summary}
Meson cloud, statistical and hybrid models describe the ratio $\bar{d}(x)$/$\bar{u}(x)$ for $x\leq 0.2$, but  fail to agree with the experimental trend to return to symmetry, or even $\bar{u}$ dominance at larger $x$. Higher statistics results from E-906/SeaQuest  \cite{SeaQuest} should determine the magnitude of the disagreement between  current theory and experiment.
\begin{theacknowledgments}
 This work has been supported in part by the RUI program of the National Science Foundation, Grant No. 0855656.
\end{theacknowledgments}


\end{document}